\documentclass[sigconf]
{acmart}
\AtBeginDocument{%
  }

\copyrightyear{2018}
\acmYear{2018}

{}
{}

\acmConference{}
{}
{}

\acmBooktitle{Woodstock '18: ACM Symposium on Neural Gaze Detection, June 03--05, 2018, Woodstock, NY}

\acmDOI{XXXXXXX.XXXXXXX}
\acmISBN{978-1-4503-XXXX-X/2018/06}

\usepackage{amssymb}
\usepackage{booktabs} 

\begin{document}

\title[Learning to Adjudicate: Verifier-Guided Multi-Agent Reasoning for E-Commerce Relevance]{Learning to Adjudicate: Verifier-Guided Multi-Agent Reasoning for E-Commerce Relevance}


\author{Shi Shu}
\authornote{Both authors contributed equally to this research.}
\authornote{Work done during internship at Alibaba International Digital Commerce Group.}
\email{shus25@mails.tsinghua.edu.cn}
\orcid{0009-0005-4881-8279}
\affiliation{%
  \institution{Tsinghua Shenzhen International Graduate School, Tsinghua University}
  \city{Shenzhen}
  \country{China}
}
\author{Hao Chen}
\authornotemark[1]
\email{ryan.ch@alibaba-inc.com}
\orcid{0009-0008-9029-8571}
\author{Baopu Qiu}
\email{qiubaopu.qbp@alibaba-inc.com}
\orcid{0009-0003-6872-4312}
\affiliation{%
  \institution{Alibaba International Digital Commerce Group}
  \city{Hangzhou}
  \country{China}
}

\author{Chao Wei}
\orcid{0009-0006-2417-6029}
\authornote{Corresponding author.}
\email{weichao.wc@alibaba-inc.com}
\affiliation{%
  \institution{Alibaba International Digital Commerce Group}
  \city{Hangzhou}
  \country{China}
}

\author{Weiru Zhang}
\orcid{0009-0007-8496-3638}
\email{weiru.zwr@alibaba-inc.com}
\affiliation{%
  \institution{Alibaba International Digital Commerce Group}
  \city{Hangzhou}
  \country{China}
}
\author{Xiu li}
\orcid{0000-0003-0403-1923}
\authornotemark[3]
\email{li.xiu@sz.tsinghua.edu.cn}
\affiliation{%
  \institution{Tsinghua Shenzhen International Graduate School, Tsinghua University}
  \city{Shenzhen}
  \country{China}
}
\author{Xiaoyi Zeng}
\orcid{0000-0002-3742-4910}
\email{yuanhan@taobao.com}
\affiliation{%
  \institution{Alibaba International Digital Commerce Group}
  \city{Hangzhou}
  \country{China}
}

\renewcommand{\shortauthors}{Shi Shu et al.}

\begin{abstract}
Relevance modeling is fundamental to e-commerce search, determining how well retrieved products satisfy the needs expressed in a user's query. Accurate judgments often require interpreting user intent, comparing product attributes, and applying business-specific matching criteria. Current methods based on large language models (LLMs) leverage their contextual understanding and reasoning capabilities, with research primarily focusing on incorporating domain-specific context and enhancing reasoning performance. Despite these advances, difficult query-product pairs remain challenging when evidence from different perspectives supports conflicting conclusions. In such cases, a single reasoning pass may fail to reconcile these conflicts and ensure that decisive evidence guides the final judgment.

To address this challenge, we propose Collaborative
Adjudication with Debate and Replay Experience (CADRE),
a framework for evidence-based adjudication in
e-commerce relevance assessment. CADRE organizes three expert agents sharing a frozen backbone to analyze
product attributes, user intent, and business rules in
separate contexts, while a trainable Verifier assesses
their supporting evidence against the applicable matching
criteria to determine relevance. Under expert disagreement, a Verifier-guided debate elicits
targeted clarification and revision of disputed
interpretations before final adjudication. We optimize
the Verifier through multi-turn group relative policy
optimization and adapt successful-experience replay
so that trajectories yielding correct terminal labels
can be reused alongside fresh rollouts in subsequent
updates. For efficient online serving, we further use
CADRE as an offline teacher to distill its relevance
judgments into a lightweight student model. Offline evaluations and online A/B tests on an e-commerce search advertising platform demonstrate that CADRE improves relevance classification and key business and relevance satisfaction metrics.
\end{abstract}

\ccsdesc[500]{Information systems~Relevance assessment}
\ccsdesc[500]{Information systems~Similarity measures}
\ccsdesc[500]{Information systems~Language models}
\ccsdesc[300]{Computing methodologies~Natural language processing}
\ccsdesc[300]{Information systems~Sponsored search advertising}

\keywords{Relevance Modeling, Large Language Model, Multi-Agent Debate, Reinforcement Learning, Experience Replay, E-Commerce Search}


\maketitle
\fancyhead[LE,RO]{}
\section{Introduction}

Relevance modeling is a cornerstone of e-commerce search,
as accurate query--product matching is essential to
effective product discovery and a satisfying user experience. Earlier approaches predominantly rely on BERT-style encoders to learn semantic interactions between queries and products from annotated data~\citep{kenton2019bert,10.1145/3534678.3539090}. More recently, large language models (LLMs) have been adopted to bring broader knowledge and stronger contextual reasoning to relevance judgments through prompting and fine-tuning~\citep{mehrdad2024large}. Chain-of-Thought (CoT) approaches further express these judgments as explicit reasoning steps that connect query understanding, product analysis, and the application of matching criteria~\citep{Tang_2025,Zhao_2025,dong_taosr1_2025}.

\begin{figure}[t]
    \centering
    \includegraphics[width=0.95\columnwidth]{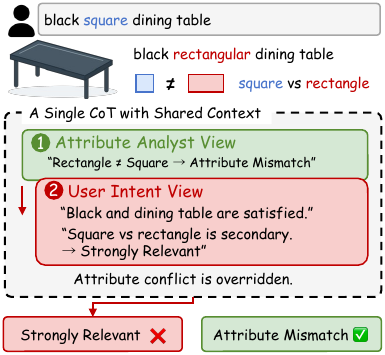}
    \caption{An illustrative case of conflict within a single CoT, where an intent-based interpretation overrides an identified attribute mismatch and yields an incorrect relevance judgment.}
    \Description{For a square-table query and a rectangular product, an initial Attribute Issue judgment is overridden by an intent view favoring Strongly Relevant.}
    \label{fig:single_cot_conflict}
\end{figure}

However, difficult query--product pairs often involve conflicting evidence from user intent, product attributes, and business rules, requiring the model to determine which evidence is decisive under the applicable matching criteria. Preference optimization methods such as DPO~\citep{rafailov2023direct} and KTO~\citep{fang2025adore}, together with RL methods such as GRPO~\citep{dong_taosr1_2025}, refine relevance reasoning through feedback on generated judgments, but resolving these conflicts remains challenging. As illustrated in Figure~\ref{fig:single_cot_conflict}, an intent-based interpretation may override a correctly identified attribute mismatch within the same reasoning context. Even when different views are generated separately, their initial reports may leave unresolved why a particular attribute should be decisive or whether a matching rule applies, requiring further examination before a final judgment can be justified. Learning from difficult cases also poses a training challenge, since trajectories yielding correct relevance labels may be scarce within a limited rollout budget and, once discovered, may not be effectively reused in subsequent updates.

\begin{figure*}[t]
    \centering
    \includegraphics[width=\textwidth]
    {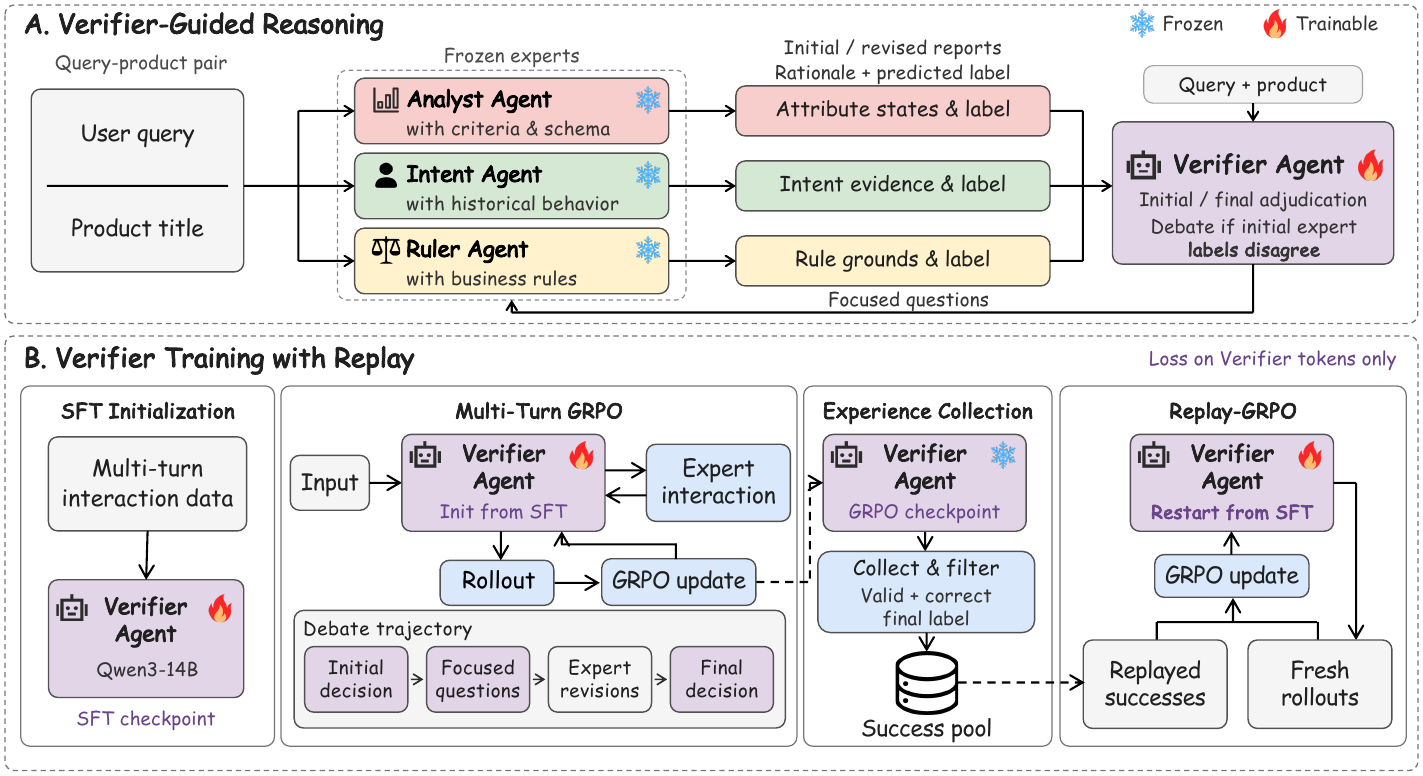}
    \caption{Overview of CADRE.
    (A) Frozen expert agents analyze query--product pairs with
    complementary context, while the Verifier integrates their
    reports and selectively initiates debate when their initial
    labels disagree.
    (B) Verifier training combines SFT initialization, multi-turn
    GRPO, successful-experience collection, and Replay-GRPO to
    reuse scarce successful trajectories and reduce repeated
    exploration.}
    \label{fig:overall}
    \Description{A two-panel architecture diagram. The upper panel
    shows expert reports, Verifier adjudication, and selective
    debate. The lower panel shows Verifier training and the
    collection and reuse of successful trajectories, with the
    expert agents remaining frozen.}
\end{figure*}

To address these challenges, we propose Collaborative
Adjudication with Debate and Replay Experience (CADRE),
a framework for evidence-based adjudication in
e-commerce relevance assessment. The framework comprises
three complementary components.

\begin{enumerate}

\item \textbf{Multi-agent analysis with a dedicated Verifier}
assigns product analysis, user intent understanding,
and business-rule interpretation to agents with
separate contexts over a shared frozen backbone.
A trainable Verifier compares their analyses and
relevance judgments to assess the supporting evidence
and determine which matching criteria apply to
the query--product pair.

\item \textbf{Verifier-guided debate} selectively extends
adjudication based on disagreement among the experts'
initial labels. For selected inputs, the Verifier asks
focused questions to clarify disputed evidence and
matching criteria, then integrates the revised expert
reports into a final judgment.

\item \textbf{Successful-experience replay} builds on
Reinforcement Learning with Experience Replay
(RLEP)~\citep{zhang2025rlepreinforcementlearningexperience}.
We adapt this approach to multi-turn Verifier training
by retaining trajectories with correct terminal labels
and their original interaction contexts, and replaying
them alongside fresh rollouts in subsequent updates.

\end{enumerate}

For efficient online deployment, we distill CADRE's
reasoning into a lightweight student that directly
scores query--product pairs.
On an in-house multilingual relevance dataset, CADRE
improves accuracy and Macro-F1 by 3.73 and 13.64
percentage points, respectively, over LLM-Replay-GRPO
with additional context. In a 7-day online A/B test, the distilled student
achieves relative gains of 1.69\% in revenue per mille
(RPM) and 0.76\% in click-through rate (CTR), alongside
an absolute increase of 0.3 percentage points in relevance
satisfaction score (RS), over the existing production
baseline.

\section{Related Works}
Early relevance models rely on lexical matching, with TF-IDF~\citep{ramos2003using} and BM25~\citep{robertson2009probabilistic} scoring query--document pairs through term statistics. Although computationally efficient, these methods have limited ability to capture semantic associations beyond lexical overlap. Neural approaches such as DSSM~\citep{10.1145/2505515.2505665} learn dense representations that bring semantically related queries and documents closer in a shared space. Pre-trained encoders such as BERT~\citep{kenton2019bert} further introduce contextual representations for semantic matching, while ColBERT~\citep{khattab2020colbert} and Poly-encoders~\citep{humeaupoly} explore different trade-offs between interaction quality and computational cost. These advances improve semantic matching, but complex product requirements and implicit shopping intent motivate methods that can incorporate broader knowledge and reason about the basis of a relevance judgment.

Recent studies therefore explore LLMs for relevance
assessment, drawing on their broad knowledge and
contextual reasoning capabilities~\citep{mehrdad2024large,wang2024improvingpinterestsearchrelevance}.
Subsequent work further grounds these capabilities
in domain-specific knowledge and matching criteria
to address the requirements of e-commerce relevance. An early effort in this direction, behavior-augmented prompting~\citep{chen2025towards}, supplies behavioral context for relevance prediction. More recent work further explores domain knowledge and matching criteria, with MPCoT~\citep{10.1145/3774904.3792812} incorporating business rules alongside intent and attribute analysis, while K-CARE~\citep{yifei2026kcareknowledgedrivensymmetricalcontextual} addresses missing domain context through behavioral anchors and expert-curated prototypes. Together, these works show the value of supplying domain context that may be absent from, or insufficiently elicited by, the model's parametric knowledge. 

Another line of research seeks to enhance reasoning quality and alignment with matching rubrics through more informative supervision. LREF~\citep{Tang_2025} and ADORE~\citep{fang2025adore} use preference optimization to calibrate relevance judgments, while SHE~\citep{jiao2026shestepwisehybridexamination}, TaoSR-AGRL~\citep{10.1145/3774904.3792799}, and LORE~\citep{lu2026lorelargegenerativemodel} develop stepwise rewards, rule-aware feedback, and progressive reasoning supervision, respectively. SHE organizes relevance assessment into five stages, and TaoSR-AGRL follows a predefined category--attribute--relevance pattern. Prior work also addresses the reliability of relevance reasoning and its alignment with final predictions. TaoSR1~\citep{dong_taosr1_2025} combines preference optimization with RL to mitigate reasoning errors and inconsistencies between reasoning and final answers. More recently, Graph-GRPO~\citep{che2026graphgrpo} models relevance reasoning as a dependency graph and propagates reward signals across dependent reasoning components, enabling fine-grained credit assignment beyond uniform sequence-level feedback. These methods strengthen the supervision of relevance
reasoning, while leaving room to investigate how
disagreements among complementary analyses can guide
targeted re-examination of evidence and matching
criteria before final adjudication.

Effective RL training also depends on obtaining and reusing informative trajectories within a limited rollout budget. In relevance modeling, TaoSR1 and SHE use difficulty-based filtering to improve the informativeness of training groups~\citep{dong_taosr1_2025,jiao2026shestepwisehybridexamination}, while TaoSR-AGRL introduces targeted ground-truth guidance to regenerate trajectories for low-reward cases~\citep{10.1145/3774904.3792799}. These strategies improve the utility of newly sampled trajectories, while reusing previously discovered successes offers a complementary direction. In the broader context of LLM reasoning, RLEP~\citep{zhang2025rlepreinforcementlearningexperience} collects successful trajectories from an RL-trained policy and combines them with fresh rollouts during a subsequent training run. These studies highlight the complementary roles of discovering useful trajectories and retaining successful experiences, providing a basis for investigating experience reuse in multi-turn relevance assessment.

Knowledge distillation provides a practical route for deploying LLM-based relevance models under online latency constraints. Recent work transfers supervision from reasoning
LLMs to lightweight BERT models~\citep{agrawal-etal-2025-rationale,Zhao_2025,10.1145/3774904.3792801}, while MPCoT~\citep{10.1145/3774904.3792812} introduces latent reasoning distillation and Graph-GRPO~\citep{che2026graphgrpo} uses graph-node-based multi-head distillation to supervise compact students with structured reasoning signals. These studies establish knowledge distillation as a practical connection between reasoning-intensive offline teachers and
efficient online relevance models.


\section{Methods}
\subsection{Relevance Task Definition}
As shown in Figure~\ref{fig:overall}, we propose Collaborative Adjudication with Debate and Replay Experience (CADRE), a framework that combines evidence adjudication with the reuse of successful reasoning attempts. We first define the e-commerce relevance task with a multi-agent setting, then introduce the multi-agent analysis generation protocol and detail how we optimize the Verifier via multi-turn agentic GRPO and experience replay. 
Finally, we distill CADRE's reasoning into a lightweight student for efficient online serving.

In e-commerce search, relevance modeling assesses how well
a product satisfies the need expressed in a user's query
under the applicable matching criteria. Given a query $q$
and a product title $p$, we denote the input pair by
$x=(q,p)$ and formulate relevance prediction as multi-class
classification over a fine-grained label space $\mathcal{Y}$
containing seven relevance categories, detailed in
Section~\ref{sec:dataset}. We write $y^{*}\in\mathcal{Y}$
for the gold label and $\hat{y}$ for a prediction.

We extend the standard single-model formulation to three
expert agents indexed by
\begin{equation}
\mathcal{K}
= \{\mathrm{Ana.},\ \mathrm{Int.},\ \mathrm{Rule}\}.
\end{equation}
Each expert receives the query--product pair and a
role-specific input context $c_k$, and produces an analysis
$r_k$ and a predicted label $\hat{y}_k\in\mathcal{Y}$:
\begin{equation}
a_k:\ (x,c_k)\ \longmapsto\ (r_k,\hat{y}_k).
\end{equation}
We refer to the analysis and predicted label
$(r_k,\hat{y}_k)$ collectively as an expert report. Here, $c_k$ denotes the context supplied to an expert
before analysis, while $r_k$ contains its reasoning
and intermediate findings derived from $(x,c_k)$.
The contexts provide role-specific criteria or
historical interaction evidence, allowing the experts
to examine complementary aspects of the same
query--product pair within a shared label space.
The context construction and analysis procedure
for each role are detailed in
Section~\ref{sec:analysis-adjudication}.

All experts share a frozen backbone model $\mathcal{M}$,
with specialization induced by their role prompts and input
contexts. The Verifier agent $V_{\theta}$, with trainable
parameters $\theta$, receives the query--product pair and
the expert reports to produce an adjudication rationale $R$
and a relevance label $\hat{y}$:
\begin{equation}
V_{\theta}:
\bigl(x,\{(r_k,\hat{y}_k)\}_{k\in\mathcal{K}}\bigr)
\ \longmapsto\ (R,\hat{y}).
\end{equation}

\subsection{Verifier-Guided Multi-Agent Reasoning}
\label{sec:multi-agent-reasoning}

Our framework organizes relevance assessment around independent
expert analyses and a dedicated Verifier. The experts examine
complementary evidence in separate contexts, while the Verifier
evaluates their reports to determine the relevance judgment.
For inputs selected by a predefined disagreement criterion,
the Verifier initiates a focused debate to obtain clarification
and revised analyses before making the final decision.

\subsubsection{Multi-Agent Analysis and Adjudication}
\label{sec:analysis-adjudication}
Prior work has leveraged diverse sources of contextual
information to support e-commerce relevance
judgments~\citep{10.1145/3774904.3792812,yifei2026kcareknowledgedrivensymmetricalcontextual}.
We build on this idea by assigning complementary analytical
roles to three expert agents, each analyzing the query-product
pair $x=(q,p)$ in a separate context over a shared frozen backbone.
Role-specific prompts and inputs induce specialization while holding model capacity fixed. Each expert develops its analysis without observing
the other experts' reports, leaving cross-perspective
comparison to the Verifier. We freeze the expert backbone
to reduce optimization overhead and retain its general-purpose
response policy, while concentrating task-specific learning
on the Verifier's evidence adjudication and targeted questioning.

The \textbf{Analyst Agent} receives
$c_{\mathrm{Ana}}$, comprising structured comparison
criteria and an attribute schema adapted from
\citet{Zhao_2025}, covering fourteen dimensions including
category, brand, and material.
It first generates attribute-level matching results
$e_{\mathrm{Ana}}$ and then reasons over them to
determine relevance.
Each attribute is assessed as a match, mismatch, or
unspecified; a mismatch requires explicitly conflicting
values on both sides, while an attribute specified on
only one side is marked as unspecified and supports
neither agreement nor conflict.
These comparison states describe attribute-level
evidence, while the final relevance judgment also
depends on the applicable matching criteria.
The \textbf{Ruler Agent} receives
$c_{\mathrm{Rule}}$, containing the original text of
the platform's relevance guidelines concerning
categories, brands, model designations, attributes,
main-product versus accessory relations, and multi-SKU
listings, including precedence rules for overlapping mismatch conditions.
It first identifies the applicable clauses
$e_{\mathrm{Rule}}$ and then derives its judgment
by reasoning about the pair under those clauses.
As the only expert receiving historical behavioral
context, the \textbf{Intent Agent} receives
$c_{\mathrm{Int}}=e_{\mathrm{Int}}$, comprising the
top-5 historical search terms associated with the
candidate product and the top-5 product titles clicked
in historical searches for the current query, to infer
the shopping need expressed by the query and assess
the candidate product's functional substitutability.

Each expert $k\in\mathcal{K}$ produces a first-round report
containing its analysis $r_k^{(1)}$ and a predicted label
$\hat{y}_k^{(1)}$. The trainable \textbf{Verifier Agent}
receives the query--product pair and all three reports,
examines the evidence supporting each interpretation,
and assesses the applicability of the matching criteria
invoked by the experts. It then produces an initial
adjudication rationale $R^{(1)}$ and a label
$\hat{y}_V^{(1)}$. The Verifier retains independent decision
authority, so its judgment may differ from the experts'
shared label even when all three agree.
The Verifier can thus retain useful observations from expert
reports even when all three initial predictions are incorrect.

\subsubsection{Selective Verifier-Guided Debate}

To control the additional cost of expert interaction, we
initiate debate if and only if the three experts'
first-round relevance labels are not unanimous, covering
both a two-to-one label split and three distinct labels.
When all three experts agree, the Verifier's initial
adjudication serves as the final prediction, even if it
differs from their shared label. For inputs selected for
debate, the Verifier examines the conflicting reports and
generates one focused question for each expert to clarify
the evidence or matching criteria underlying its judgment.

Each expert receives its own question together with its first-round report and returns a revised analysis and label $(r_k^{(2)},\hat{y}_k^{(2)})$. The questions direct attention to disputed interpretations, such as whether an attribute mismatch is decisive or whether a business rule applies to the current pair. Experts respond independently, without observing one another's reports or communicating directly, while the Verifier coordinates the examination of disagreements across their analyses. After this single round of questions and responses, the Verifier integrates the initial reports and subsequent feedback to produce the final adjudication rationale $R^{(2)}$ and label $\hat{y}_V^{(2)}$. This bounded protocol allows disputed evidence to be re-examined while concentrating additional interaction on cases with expert disagreement.

\subsection{Verifier Training with Experience Replay}
\label{sec:verifier-training}

We train the Verifier to perform the adjudication and
questioning described in
Section~\ref{sec:multi-agent-reasoning}, while keeping
the shared backbone Qwen3-14B of the three expert agents frozen.
As illustrated in Figure~\ref{fig:overall}, training
proceeds through supervised initialization, on-policy
optimization, experience collection, and optimization
with replay. Supervised fine-tuning (SFT) on multi-turn
interaction data produces an initial checkpoint
$\theta_{\mathrm{SFT}}$, which is subsequently optimized
with group relative policy optimization (GRPO).
The resulting checkpoint $\theta_{\mathrm{GRPO}}$ is
used to collect successful interaction trajectories.
Replay-GRPO then restarts from $\theta_{\mathrm{SFT}}$
and combines fresh rollouts with the collected experiences.

The original dataset provides relevance labels but
does not contain annotated adjudications or questions
for Verifier training. Using only the training split,
we construct synthetic supervision by combining expert
reports generated by the frozen Qwen3-14B backbone
with Verifier outputs generated through the Qwen3.8-Max
API. The expert models produce their initial reports
and subsequent revisions, while Qwen3.8-Max generates
the Verifier's adjudications and expert-specific
questions, with the annotated relevance label supplied
as additional guidance in its prompt. During SFT,
this label-bearing guidance is removed from the
Verifier's input, while the expert reports remain
part of the interaction context and the generated
Verifier responses serve as supervision targets.
The resulting checkpoint $\theta_{\mathrm{SFT}}$
initializes subsequent multi-turn policy optimization.

\subsubsection{Multi-Turn Policy Optimization}
\label{sec:multi-turn-optimization}

We formulate Verifier training as a single-agent reinforcement learning problem in which the frozen experts and message templates constitute the interaction environment. The Verifier generates adjudications and focused questions, while the environment supplies instructions and expert responses. For an input that satisfies the predefined debate-triggering
criterion, we represent the complete interaction as
$\tau = (u_1,o_1,u_2,o_2,u_3,o_3)\label{eq:verifier-trajectory}$, where $u_1$ contains the query--product pair and the initial expert
reports, $u_2$ provides the instruction for generating focused
questions, and $u_3$ contains the revised expert reports. The
Verifier outputs $o_1$, $o_2$, and $o_3$ correspond to the initial
adjudication, the three expert-specific questions, and the final
adjudication, respectively. When the input does not satisfy the
debate-triggering criterion, the trajectory consists only of
$(u_1,o_1)$, and the Verifier's initial adjudication serves as the
final output. In either branch, the Verifier determines its judgment
independently of whether the expert labels agree.

We assign a policy-loss mask $m=1$ to Verifier-generated
tokens and $m=0$ to environment messages and expert
responses, which remain part of the conditioning context.
Table~\ref{tab:training_turns} in
Appendix~\ref{app:interaction-protocol} details the
interaction steps and their masks.
Starting from $\theta_{\mathrm{SFT}}$, we sample a group
of complete trajectories for each training input and
assign each trajectory a scalar reward $r(\tau)$,
defined in Section~\ref{sec:composite-reward}.
GRPO computes group-relative advantages from these
rewards and applies the clipped policy objective to
Verifier-generated tokens under their corresponding
interaction histories.

\subsubsection{Reward Design}
\label{sec:composite-reward}

We assign each complete Verifier trajectory $\tau$
a scalar reward combining adjudication quality
and debate quality:
\begin{equation}
r(\tau)
=
r_{\mathrm{adj}}
+
\operatorname{Gate}_{\mathrm{debate}}
r_{\mathrm{shape}},
\label{eq:composite-reward}
\end{equation}
where $\operatorname{Gate}_{\mathrm{debate}}=1$
when the three experts' first-round labels are
not unanimous and is zero otherwise.

\noindent\textbf{Adjudication quality.}
The reward $r_{\mathrm{adj}}$ combines format
validity $r_{\mathrm{fmt}}$, hierarchical label
accuracy $r_{\mathrm{res}}$, and reasoning checks
$r_{\mathrm{reason}}$.
The terminal prediction is the Verifier's initial
judgment without debate and its final judgment
after debate.
Outcome supervision assigns full credit to an
exact label match and partial credit to a correct
relevant/irrelevant superclass, while reasoning
checks assess rationale--label consistency and
adjudication completeness.
Outcome and reasoning rewards are enabled only
when the terminal label is valid.

\noindent\textbf{Debate quality.}
The reward $r_{\mathrm{shape}}$ evaluates the
initial judgment, question quality, and expert
revisions.
It rewards a correct initial decision and uses
question length and references to input tokens
as heuristics for question specificity, while
penalizing malformed questions and disclosure
of the Verifier's own judgment.
Expert revisions receive greater credit for
correcting an error than for preserving a correct
prediction, with penalties for regressions.
These revisions are evaluated against the gold
label, allowing experts to disagree with the
Verifier's initial judgment.
Detailed scoring rules and coefficients are
provided in Appendix~\ref{app:reward-details}.

\subsubsection{Successful-Experience Replay}
\label{sec:verified-replay}

Although the reward provides supervision for adjudication
and questioning, a limited rollout budget may yield no
successful trajectory for a difficult input, while
occasional successes are not automatically retained
for subsequent updates.
Following RLEP~\citep{zhang2025rlepreinforcementlearningexperience},
we adapt successful-experience replay to the Verifier's
interaction with frozen experts through a per-input
experience pool $\mathcal{B}(x)$.
The pool retains trajectories with correct terminal
labels together with their original interaction contexts,
allowing successful adjudication and questioning attempts
to be reused alongside fresh rollouts during Verifier
training.

After on-policy GRPO, the checkpoint
$\theta_{\mathrm{GRPO}}$ generates 16 complete
trajectories per input using the same Verifier--expert
interaction protocol as in training.
We define success by a parseable terminal prediction
that exactly matches the gold label, so every retained
trajectory satisfies
\begin{equation}
\tau\in\mathcal{B}(x)
\ \Longrightarrow\
r_{\mathrm{res}}(\tau)=1
\ \land\
r_{\mathrm{fmt}}(\tau)=1.
\label{eq:replay-admission}
\end{equation}
Among the trajectories satisfying these conditions,
we retain at most eight per input, ranked by their
total reward.
The auxiliary reward terms therefore determine the
ranking of successful candidates, while admission
requires an exactly correct terminal label.
Each stored trajectory preserves its original token
IDs and policy-loss mask, maintaining the alignment
between Verifier outputs and interleaved environment
messages.

Replay-GRPO restarts from $\theta_{\mathrm{SFT}}$,
the same initialization used for on-policy GRPO.
For each input, we combine $G$ fresh rollouts with
$M$ trajectories sampled from $\mathcal{B}(x)$, with empty pools contributing none. Excess replay rows are randomly subsampled, while batch deficits are filled by resampling with replacement within the original input groups; if the budget cannot be met, the update uses fresh-only GRPO. All trajectories are scored through a unified reward pipeline, and group-relative advantages are computed over the actual group:
\begin{equation}
A_i=
\frac{r_i-\operatorname{mean}(r_1,\ldots,r_{n_x})}
{\operatorname{std}(r_1,\ldots,r_{n_x})+\epsilon},
\qquad i=1,\ldots,n_x.
\label{eq:replay-advantage}
\end{equation}
Here, $n_x$ denotes the actual number of trajectories
for input $x$ in the update, and $r_i$ is the reward
of the $i$-th trajectory.
We set $\epsilon=10^{-6}$ to prevent division by zero,
yielding zero advantages for equal-reward groups.
Fresh and replayed trajectories enter the same clipped
GRPO objective, with the policy loss applied only to
Verifier-generated tokens.
Old log-probabilities are recomputed using the update-start
actor, without importance correction for historical
generating policies.
Reusing scarce successful trajectories allows each
discovered success to contribute to multiple updates,
reducing reliance on repeated exploration to recover
successful interactions.

\subsection{Knowledge Distillation}
\label{sec:kd}

CADRE adjudicates each pair through multiple expert generations
and a Verifier, which is too costly for low-latency online serving.
We therefore train a lightweight \textbf{student} $g_\phi$:
an mBERT encoder with a seven-way classification head that
maps $x=(q,p)$ directly to $\hat{y}\in\mathcal{Y}$.
Let $p_\phi(\cdot\mid x)$ denote the student's predicted
distribution over $\mathcal{Y}$.
Gold relevance labels supervise classification, while CADRE
serves as an offline \textbf{teacher} providing CoTs as
auxiliary supervision.
Following prior work on reasoning
distillation~\citep{10.1145/3774904.3792801}, we transfer
this reasoning signal into the student's representation
rather than generating it at inference time.

Let $c$ denote the teacher's CoT for input $x$.
We use a frozen BGE-M3 encoder~\citep{chen2024bgem3embeddingmultilingualmultifunctionality}
as $f_{\text{emb}}$ to obtain a dense target representation
$e_{\text{cot}}=f_{\text{emb}}(c)$.
The student encodes $x$ with mBERT and takes
its $\mathbf{h}_{[\mathrm{CLS}]}$ representation, which a
projection layer maps to $e_r=g_\phi^{\text{proj}}(x)$.
In addition to the standard classification loss
$\mathcal{L}_{\text{CE}}(p_\phi(\cdot\mid x), y^{*})$,
we align the student's representation with the teacher's
reasoning through a cosine distance:
\begin{equation}
\label{eq:kd-loss}
    \mathcal{L}
    = \mathcal{L}_{\text{CE}}(p_\phi(\cdot\mid x), y^{*})
    + \lambda\, \mathcal{L}_{\cos}(e_r, e_{\text{cot}}), \qquad
    \mathcal{L}_{\cos}
    = 1 - \frac{e_r \cdot e_{\text{cot}}}
    {\lVert e_r\rVert\, \lVert e_{\text{cot}}\rVert},
\end{equation}
where $\lambda$ balances reasoning distillation against
gold-label supervision. The CoT branch and $f_{\text{emb}}$
are used only during training; at inference the student
scores $x$ alone, with no multi-agent interaction and
no reasoning generation.

\section{Experiments}
\label{sec:experiments}
\begin{table*}[t]
\centering
\caption{Teacher and student performance on the in-house
dataset, with the best results in each group shown in
bold and the second-best results underlined.}
\label{tab:teacher-performance}
\begin{tabular}{lcccccc}
\toprule
\textbf{Method} &
\multicolumn{3}{c}{\textbf{Seven-Class Classification}} &
\multicolumn{3}{c}{\textbf{Irrelevant Classification}} \\
\cmidrule(lr){2-4} \cmidrule(lr){5-7}
& \textbf{Accuracy} & \textbf{Macro-F1} & \textbf{Weighted-F1}
& \textbf{Precision} & \textbf{Recall} & \textbf{F1} \\
\midrule

\multicolumn{7}{l}{\textbf{Teacher models (Qwen3-14B)}} \\
\addlinespace

\multicolumn{7}{l}{
     \textit{Vanilla single model}} \\
\quad LLM-Base
& 0.5408 & 0.4799 & 0.6092 
& 0.7511 & 0.7318 & 0.7414 \\


\quad LLM-SFT
& 0.7278 & 0.6305 & 0.7247 
& 0.8507 & 0.7916 & 0.8201 \\

\quad LLM-GRPO
    & 0.7749 & 0.6173 & 0.7720
    & 0.8562 & 0.8924 & 0.8739 \\
\quad LLM-Replay-GRPO
    & 0.7906 & 0.6369 & 0.7877
    & 0.8615 & \textbf{0.9007} & 0.8806 \\
\addlinespace

\multicolumn{7}{l}{
    \textit{Single model with additional context}} \\
\quad LLM-Base
& 0.6431 & 0.4964 & 0.6426 
& 0.7986 & 0.7200 & 0.7573 \\

\quad LLM-SFT
& 0.7350 & 0.6380 & 0.7320 
& 0.8638 & 0.7945 & 0.8277 \\

\quad LLM-GRPO
    & 0.7823 & 0.6268 & 0.7805
    & 0.8593 & 0.8951 & 0.8768 \\
\quad LLM-Replay-GRPO
    & 0.7961 & 0.6402 & 0.7915
    & 0.8629 & \underline{0.8994} & 0.8808 \\
\addlinespace

\multicolumn{7}{l}{
    \textit{Multi-agent with additional context; without debate}} \\
\quad Verifier-Base
 & 0.6369 & 0.5402 & 0.6685 
 & 0.7449 & 0.8974 & 0.8141 \\
\quad Expert Oracle
    & 0.8051 & -- & --
    & -- & -- & -- \\
\quad Verifier-SFT
    & 0.8020 & 0.7351 & 0.8045
    & 0.8705 & 0.8868 & 0.8786 \\
\quad Verifier-GRPO
    & 0.8239 & 0.7643 & 0.8230
    & 0.8939 & 0.8861 & 0.8900 \\
\quad Verifier-Replay-GRPO
    & \underline{0.8311}
    & \underline{0.7729}
    & \underline{0.8296}
    & 0.9028 & 0.8815 & \underline{0.8920} \\
\addlinespace

\multicolumn{7}{l}{
    \textit{Multi-agent with additional context; with debate}} \\
\quad Verifier-SFT
    & 0.7956 & 0.7136 & 0.7961
    & 0.8712 & 0.8791 & 0.8751 \\
\quad Verifier-GRPO
    & 0.8275 & 0.7649 & 0.8254
    & \textbf{0.9050} & 0.8729 & 0.8887 \\
\quad \textbf{CADRE (Ours)}
    & \textbf{0.8334}
    & \textbf{0.7766}
    & \textbf{0.8321}
    & \underline{0.9035} & 0.8840 & \textbf{0.8936} \\
    
\addlinespace

\multicolumn{7}{l}{
    \textbf{Student models (mBERT; CoT teacher: CADRE)}} \\
\addlinespace
\quad Student-Label-Only
& 0.7456 & 0.6726 & 0.7292 
& \textbf{0.8679} & 0.7387 & 0.7981 \\

\quad Student-KD (Experts CoTs)
& 0.7694 & 0.7229 & 0.7663 
& 0.8457 & \underline{0.8238} & 0.8346 \\
\quad Student-KD (Verifier CoT)
& \underline{0.7795} & \underline{0.7306} & \underline{0.7725} 
& \underline{0.8609} & 0.8137 & \underline{0.8366} \\

\quad Student-KD (Experts CoTs + Verifier CoT)
& \textbf{0.7809} & \textbf{0.7391} & \textbf{0.7797 }
& {0.8461} & \textbf{0.8471} & \textbf{0.8466} \\
\bottomrule
\end{tabular}
\end{table*}

\subsection{Dataset}
\label{sec:dataset}

We evaluate CADRE on an in-house e-commerce dataset with
78,997 training and 18,710 test query--product pairs.
Each pair is labeled as \textit{Strongly Relevant},
\textit{Substitute}, or one of five irrelevant categories:
\textit{Brand Mismatch}, \textit{Category Mismatch},
\textit{Model Mismatch}, \textit{Attribute Mismatch},
and \textit{Main Product vs. Accessory}.
When multiple mismatch conditions apply to the same query-product pair, annotators assign a single label, following the most severe applicable mismatch.
To mitigate label imbalance, we use label-specific
sampling quotas targeting an approximately 1:1 ratio
between relevant and irrelevant pairs and roughly
balanced representation of the five irrelevant categories.
The training and test splits have similar label proportions.
Detailed dataset statistics are provided in
Appendix~\ref{app:dataset-statistics}.

\subsection{Baselines and Evaluation Metrics}
\subsubsection{Baselines}

All teacher-stage methods in
Table~\ref{tab:teacher-performance} use Qwen3-14B
as the backbone.
Additional context
$C=(c_{\mathrm{Ana}},e_{\mathrm{Int}},c_{\mathrm{Rule}})$
comprises the attribute schema and comparison criteria,
historical search and click context associated with
the query and product, and the full text of the
platform relevance guidelines, respectively.
We evaluate single-model baselines with and without $C$.
The context-augmented variants directly process the same
source information distributed across the three experts.
\textbf{LLM-Base} denotes the off-the-shelf checkpoint
without task-specific training.
\textbf{LLM-SFT} applies supervised fine-tuning using
relevance reasoning and predictions generated by
Qwen3.8-Max under gold-label guidance.
This follows the same supervision-generation approach
as Verifier-SFT, with the prompt format adapted to
single-model relevance prediction.
\textbf{LLM-GRPO} and \textbf{LLM-Replay-GRPO}
apply GRPO and Replay-GRPO, respectively, starting
from the corresponding LLM-SFT checkpoint.

For multi-agent methods, \textbf{Verifier-Base} uses the
off-the-shelf checkpoint to adjudicate expert reports.
\textbf{Verifier-SFT} applies supervised fine-tuning,
while \textbf{Verifier-GRPO} and
\textbf{Verifier-Replay-GRPO} apply the corresponding
RL methods after supervised initialization.
We compare variants with and without selective debate,
with \textbf{CADRE} denoting Verifier-Replay-GRPO
with debate. Using the same initial expert reports as CADRE,
\textbf{Expert Oracle} measures the maximum Accuracy
achievable by selecting among the three experts'
initial labels, counting an input as correctly
classified whenever at least one expert predicts
its gold seven-class label.
Only Accuracy is reported because the oracle leaves
the selected label unspecified when all experts are
incorrect, preventing the other metrics from being
uniquely determined.

For fair offline comparisons, all student variants use
mBERT and the same 78,997-pair training split used for
teacher training. Gold labels supervise classification,
with CADRE providing CoTs for the distillation variants.
\textbf{Student-Label-Only} uses gold labels alone.
\textbf{Student-KD (Experts CoTs)} additionally uses
concatenated CoTs from the three expert agents, while
\textbf{Student-KD (Verifier CoT)} uses the Verifier's
adjudication rationale.
\textbf{Student-KD (Experts CoTs + Verifier CoT)}
combines both sources of reasoning supervision.
For inputs that do not trigger debate, only
first-round CoTs are available.

\subsubsection{Evaluation Metrics}
For offline evaluation, we assess both fine-grained relevance classification and irrelevant-product identification. For seven-class classification, we report Accuracy (ACC), Macro-F1, and Weighted-F1. Since irrelevant product exposure can degrade the search experience~\citep{che2026graphgrpo}, we additionally report Precision, Recall, and F1 for irrelevant classification. We map both predicted and gold labels to a binary label space, treating \textit{Strongly Relevant} and \textit{Substitute} as negative and the remaining five categories as positive (irrelevant). These metrics assess the model's ability to identify
irrelevant products while limiting erroneous rejection
of relevant candidates.

\subsection{Implementation Details}

We conduct experiments on a single node with 8
NVIDIA H100 GPUs (80\,GB each).
All agents in CADRE use Qwen3-14B as their backbone.
The experts share a frozen model instance with separate
contexts, while only the Verifier is optimized.
We implement this interaction and masking procedure
through the agent loop in
\texttt{verl}~\citep{sheng2024hybridflow}%
\footnote{\url{https://github.com/verl-project/verl}}.
Within each experimental setting, GRPO and Replay-GRPO
start from the same SFT checkpoint and sample $G=8$
fresh trajectories per input, with Replay-GRPO
additionally incorporating $M=2$ stored successful
trajectories.
Detailed training and replay configurations are provided
in Appendix~\ref{app:training-configuration}.

\subsection{Performance}
\label{sec:performance}

\subsubsection{Overall Performance.}
As shown in Table~\ref{tab:teacher-performance}, CADRE
achieves the highest reported seven-class classification
scores, with 0.8334 ACC, 0.7766 Macro-F1, and 0.8321
Weighted-F1. Compared with LLM-Replay-GRPO with additional
context, these results represent improvements of 3.73,
13.64, and 4.06 percentage points, respectively.
CADRE also achieves the highest irrelevant-class F1
of 0.8936, indicating stronger overall performance
in identifying irrelevant query--product pairs.
Together with its fine-grained mismatch predictions,
this capability can support filtering decisions that
prioritize severe mismatches, helping reduce irrelevant
product exposure and improve the search experience.

Expert Oracle achieves an accuracy of 0.8051. CADRE exceeds this coverage by 2.83 percentage points,
indicating that the framework can correctly adjudicate
some instances where all three initial expert labels
are incorrect. This result supports CADRE's robustness
to imperfect initial expert predictions, as its final
adjudication draws on supporting evidence and any debate
feedback without requiring every expert to predict
the correct label. Appendix~\ref{app:case-studies} provides qualitative
examples of independent adjudication and correction
through Verifier-guided debate.

\subsubsection{Effectiveness of CADRE}
\label{sec:cadre-effectiveness}
With additional context, Verifier-SFT without debate
outperforms LLM-SFT by 6.70, 9.71, and 7.25 percentage
points in ACC, Macro-F1, and Weighted-F1, respectively.
Without debate, Verifier-GRPO outperforms Verifier-SFT
by 2.19, 2.92, and 1.85 percentage points in ACC, Macro-F1,
and Weighted-F1, respectively, demonstrating the benefit
of reinforcement learning for adjudicating expert reports. CADRE also outperforms Verifier-Replay-GRPO without
debate across all six reported metrics, supporting
the contribution of selective debate within the
Replay-GRPO setting.
By incorporating selective debate and successful-experience
replay, CADRE further extends the gains over the same
Verifier-SFT baseline to 3.14, 4.15, and 2.76 points,
respectively.

\subsubsection{Effectiveness of KD}

We compare mBERT students trained with different CoT components
produced by CADRE. Compared with gold-label supervision alone,
the best configuration, \textit{Experts CoTs + Verifier CoT},
improves ACC, Macro-F1, and Weighted-F1 by 3.53, 6.65,
and 5.05 percentage points, respectively, and
irrelevant-class F1 by 4.85 points.
These results support the value of the teacher's reasoning
beyond label supervision.
Among the CoT-based variants, \textit{Verifier CoT}
outperforms the concatenated \textit{Experts CoTs} across
all seven-class metrics (e.g., Macro-F1 0.7306 vs.\ 0.7229).
This suggests that the trained Verifier's adjudicated reasoning
provides more effective distillation supervision than simply
concatenating expert analyses.

\subsection{Ablation Studies}
We examine three components of CADRE, including the reward components, the
debate protocol, and the successful-experience replay. We count the initial expert reports as the
first expert-response round and each subsequent round of
revisions as an additional round. Unless otherwise specified,
the variants retain three expert inputs and use two
expert-response rounds, consisting of the initial reports and
one round of revisions following Verifier-guided questioning,
without direct communication among experts.

\subsubsection{Reward Components}
\label{sec:ablation-reward}

\begin{table}[t]
\centering
\caption{Ablation of the composite reward.}
\label{tab:reward-ablation}
\begin{tabular}{@{}lccc@{}}
\toprule
Method & ACC & Macro-F1 & Weighted-F1 \\
\midrule
\textbf{Full CADRE}       & 0.8334 & 0.7766 & 0.8321 \\
\quad w/o reasoning reward & 0.8261 & 0.7685 & 0.8241 \\
\quad w/o shaping reward  & 0.8267 & 0.7695 & 0.8254 \\
\bottomrule
\end{tabular}
\end{table}

Table~\ref{tab:reward-ablation} shows that both auxiliary reward
terms contribute to performance. Removing the reasoning reward
decreases ACC, Macro-F1, and Weighted-F1 by 0.73, 0.81, and 0.80
percentage points, respectively. Removing debate shaping produces
corresponding decreases of 0.67, 0.71, and 0.67 points. These
results support rewarding both a consistent adjudication process
and productive expert revisions in addition to final-label
correctness.

\begin{figure}[t]
    \centering
    \includegraphics[width=\columnwidth,keepaspectratio]
    {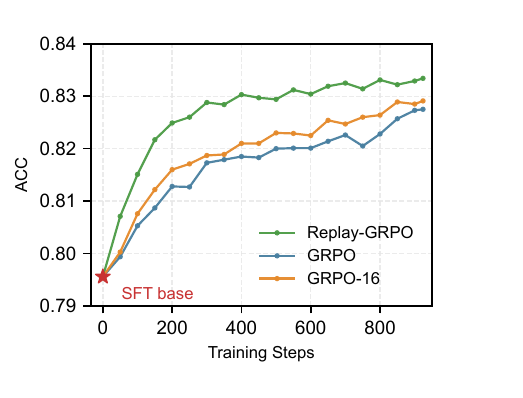}
    \caption{Test-set accuracy of GRPO, GRPO-16, and
    CADRE (Replay-GRPO) in the debate-enabled setting.}
    \Description{Test-set accuracy curves over training steps
    for GRPO, GRPO-16, and CADRE trained
    with Replay-GRPO, initialized from the same SFT checkpoint.}
    \label{fig:val_acc}
\end{figure}

\subsubsection{Debate Protocol}


\label{sec:ablation-debate}
\begin{table}[t]
\centering
\caption{Ablation of debate rounds and expert report sharing.
M-F1 and W-F1 denote Macro-F1 and Weighted-F1.}
\label{tab:debate-ablation}
\setlength{\tabcolsep}{3pt}
\begin{tabular}{@{}lccc@{}}
\toprule
Method & ACC & M-F1 & W-F1 \\
\midrule
\textbf{CADRE (2 rounds, independent)}
    & 0.8334 & 0.7766 & 0.8321 \\
3 rounds, independent
    & 0.8339 & 0.7768 & 0.8324 \\
2 rounds, initial-stage sharing
    & 0.8220 & 0.7610 & 0.8219 \\
2 rounds, debate-stage sharing
    & 0.8158 & 0.7521 & 0.8168 \\
\bottomrule
\end{tabular}
\end{table}



Table~\ref{tab:debate-ablation} examines expert report
sharing at different stages and the number of
expert-response rounds.
CADRE uses two rounds, keeping both initial analyses
and subsequent expert revisions independent. With \textit{initial-stage sharing}, experts generate
their first-round reports sequentially, and each expert
can observe the reports produced by preceding experts;
the subsequent debate retains the original
Verifier-guided revision protocol.
This variant reduces ACC, Macro-F1, and Weighted-F1
by 1.14, 1.56, and 1.02 percentage points relative
to CADRE, respectively.
This degradation is consistent with the concern
illustrated in Figure~\ref{fig:single_cot_conflict},
where one perspective overrides decisive evidence
identified from another. With \textit{debate-stage sharing}, initial analyses
remain independent, but each expert can directly
observe the other experts' analyses and predicted
labels when revising its own report.
This variant reduces the same metrics by 1.76, 2.45,
and 1.53 percentage points, respectively.
Exposure to peer judgments may encourage premature
convergence at either stage, reducing the complementary
evidence available to the Verifier.
These results support preserving independent expert
analyses and revisions.

With independent experts, extending the interaction
from two to three response rounds yields only marginal
gains of 0.05, 0.02, and 0.03 percentage points in ACC,
Macro-F1, and Weighted-F1, respectively, while requiring
an additional round of expert calls.

\subsubsection{Successful-Experience Replay}
\label{sec:ablation-replay}

As shown in Table~\ref{tab:teacher-performance},
successful-experience replay improves Verifier performance
under both interaction settings.
Without debate, Verifier-Replay-GRPO outperforms
Verifier-GRPO by 0.72, 0.86, and 0.66 percentage points
in ACC, Macro-F1, and Weighted-F1, respectively;
with debate, the corresponding gains of CADRE over
Verifier-GRPO are 0.59, 1.17, and 0.67 points.
Figure~\ref{fig:val_acc} compares accuracy on the test set
in the debate-enabled setting, with all methods
initialized from the same SFT checkpoint. These test-set evaluations were conducted retrospectively
and were not used for checkpoint selection, early stopping,
or hyperparameter tuning. By replaying successful trajectories, CADRE consistently
achieves higher accuracy than GRPO and GRPO-16, whereas
increasing fresh rollouts from 8 to 16 yields only modest
gains over standard GRPO in this setting.

\subsection{Online Evaluation}
For deployment, we train the best student on 1.64 million
human-labeled query--product pairs, using gold labels
for classification and CADRE-generated CoTs for
reasoning distillation.
We conducted a 7-day online A/B test in an e-commerce
search advertising system, allocating 10\% of user
traffic each to this student and the existing
production baseline.
During serving, predefined business rules and calibrated
thresholds map query--item relevance predictions to
discrete levels, filtering candidates below the
relevance threshold before ranking. We assess business performance using Revenue Per Mille
(RPM) and Click-Through Rate (CTR), and relevance using
Relevance Satisfaction Score (RS), defined as the
platform's traffic-weighted strong-relevance metric.
We sample results for the same queries from the two
experimental buckets under the same conditions and
submit them to human experts for relevance assessment.
The query set covers approximately 90\% of search
traffic, and RS aggregates query-level strong-relevance
rates weighted by each query's search traffic.
Compared with the existing production baseline, the
CADRE-distilled student achieves relative gains of
\textbf{1.69}\% in RPM and \textbf{0.76}\% in CTR,
alongside an absolute increase of
\textbf{0.3} percentage points in RS. These results demonstrate that the CADRE-distilled
student improves commercial performance while also
increasing the traffic-weighted strong-relevance rate.

\section{Conclusion}

In this paper, we introduced CADRE to improve evidence integration in
e-commerce relevance modeling. Its central design
separates complementary expert analysis from learned
adjudication, enabling a Verifier to weigh evidence
across perspectives and seek clarification through
selective debate. Multi-turn reinforcement learning
optimizes this adjudication process, while experience
replay preserves successful trajectories for reuse
during training. We further transfer CADRE's reasoning to a lightweight
student for online deployment.
Offline experiments demonstrate improved relevance
prediction, while online A/B testing confirms that
the distilled student improves business performance
and relevance satisfaction over the existing
production baseline. Given the current fixed-round debate protocol, future
work will explore adaptive policies for expert selection,
evidence clarification, and stopping to tailor the scope
and duration of interaction to input difficulty while
accounting for computational cost.

\bibliographystyle{ACM-Reference-Format}
\bibliography{references}

\appendix

\begin{figure*}[t]
    \centering
    \includegraphics[width=0.95\textwidth]
    {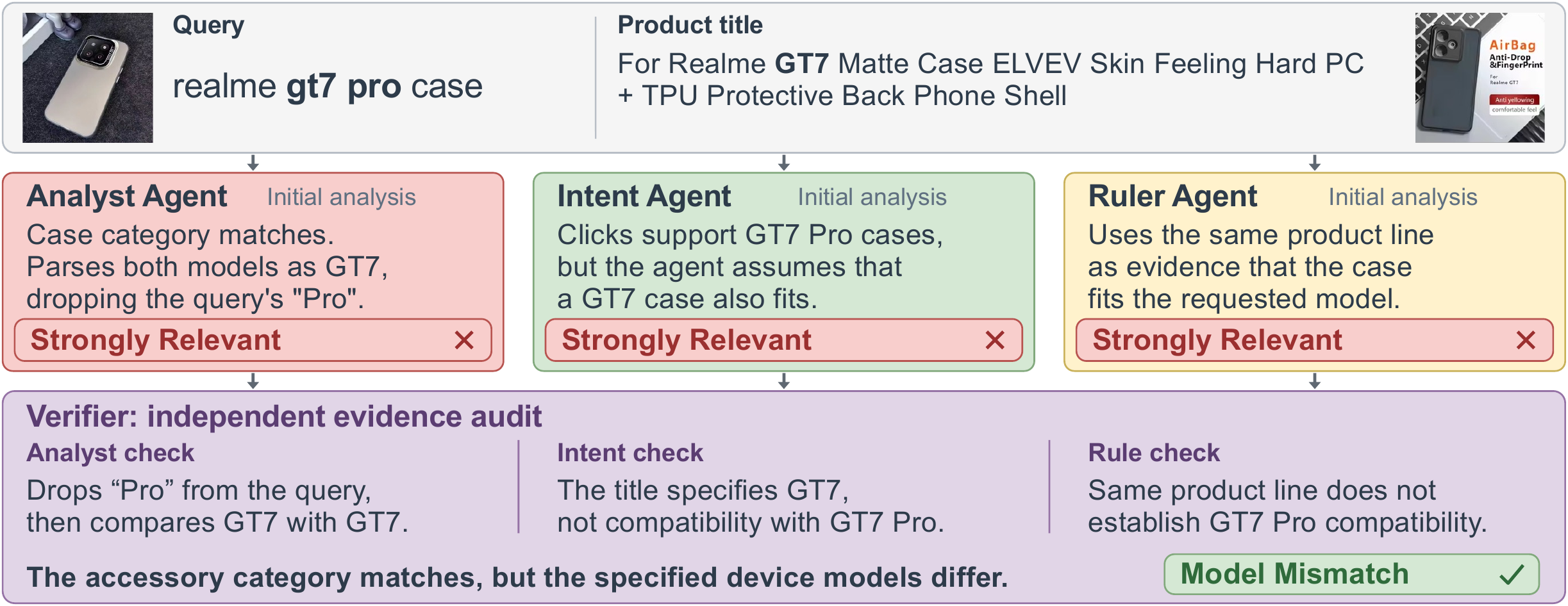}
    \caption{Correcting unanimous expert errors without debate.}
    \Description{All three experts predict Strongly Relevant.
    The Verifier identifies the GT7 Pro versus GT7 model
    mismatch and predicts Model Mismatch without debate.}
    \label{fig:case_1}
\end{figure*}

\section{Interaction Protocol and Policy-Loss Masking}
\label{app:interaction-protocol}

\begin{table}[t]
    \centering
    \caption{Debate interactions and policy-loss masks.}
    \label{tab:training_turns}
    \setlength{\tabcolsep}{3pt}
    \begin{tabular}{@{}c p{0.78\columnwidth} c@{}}
        \toprule
        \textbf{Step} & \textbf{Interaction} & $\boldsymbol{m}$ \\
        \midrule
        1 &
        \textbf{Verifier:} Initial adjudication from
        the three expert reports.
        & 1 \\

        2 &
        \textbf{Environment:} Question instruction
        with the initial judgment and expert reports.
        & 0 \\

        3 &
        \textbf{Verifier:} One focused question
        per expert.
        & 1 \\

        4 &
        \textbf{Experts:} Each revises its report
        using its own question and initial report.
        & 0 \\

        5 &
        \textbf{Environment:} Append revised reports
        to the Verifier's context.
        & 0 \\

        6 &
        \textbf{Verifier:} Final adjudication from
        the initial and revised reports.
        & 1 \\
        \bottomrule
    \end{tabular}
\end{table}

Table~\ref{tab:training_turns} summarizes the debate
interaction sequence. Only Verifier-generated tokens
contribute to the policy loss ($m=1$), while expert
outputs and environment messages are masked ($m=0$).
For inputs not selected for debate, the initial
adjudication serves as the final prediction.

\section{Dataset Statistics}
\label{app:dataset-statistics}

\begin{table}[t]
\centering
\caption{Seven-class label distribution of the full dataset.}
\label{tab:dataset-label-distribution}
\begin{tabular}{@{}lrr@{}}
\toprule
\textbf{Label} & \textbf{Count} & \textbf{Share (\%)} \\
\midrule
Strongly Relevant          & 49,955 & 51.13 \\
Substitute                 & 1,114  & 1.14  \\
Category Mismatch          & 9,326  & 9.54  \\
Brand Mismatch             & 9,135  & 9.35  \\
Model Mismatch             & 9,145  & 9.36  \\
Attribute Mismatch         & 10,454 & 10.70 \\
Main Product vs. Accessory & 8,578  & 8.78  \\
\addlinespace
\textbf{Total} & \textbf{97,707} & \textbf{100.00} \\
\bottomrule
\end{tabular}
\end{table}

Table~\ref{tab:dataset-label-distribution} reports the final
cleaned labels for 97,707 pairs, with 78,997 training and
18,710 test pairs whose class proportions closely match.
Relevant and irrelevant pairs account for 52.27\% and
47.73\%, respectively. Macro-F1 complements Accuracy and
Weighted-F1 by weighting classes equally despite their
unequal frequencies. The five most represented languages
are English, Spanish, Portuguese, Korean, and French.

\section{Implementation and Training Configuration}
\label{app:training-configuration}

The experts share a frozen Qwen3-14B instance at a dedicated
vLLM endpoint, with separate prompts, contexts, and
conversation histories, each incurring a generation cost.
The Verifier is a separate model initialized through
full-parameter SFT on multi-turn interactions using
LLaMA-Factory\footnote{\url{https://github.com/hiyouga/LlamaFactory}}.
Training runs for 3 epochs with a learning rate of
$1 \times 10^{-5}$ and a batch size of 16, using BF16
precision and ZeRO-2.

Both GRPO and Replay-GRPO use \texttt{verl}, starting from
$\theta_{\mathrm{SFT}}$ and training for 3 epochs with a
learning rate of $1 \times 10^{-6}$, a batch size of 256,
and $G=8$ fresh trajectories per input.

For replay, $\theta_{\mathrm{GRPO}}$ generates 16 trajectories
per input at temperature 1.0 using the same Verifier--expert
interaction protocol as in training.
Following Section~\ref{sec:verified-replay}, we retain at most
$K=8$ trajectories with parseable, exactly correct terminal
predictions, ranked by total reward. Replay-GRPO adds $M=2$
stored trajectories per input to the fresh rollouts.

The mBERT-base student is
trained for 3 epochs with a learning rate of $2 \times 10^{-5}$,
a batch size of 128, and $\lambda=0.1$.
For each training pair, we concatenate the CoTs generated across
all available rounds by the Analyst, Intent, and Ruler Agents
and the Verifier. Each variant includes only the CoTs from its
corresponding agents. The concatenated
text is encoded by a frozen BGE-M3 model into a single embedding
$e_{\mathrm{cot}}$. Since the concatenated CoTs fit within
BGE-M3's maximum length of 8,192 tokens, no truncation is applied. No correctness-based filtering is applied to the
CoTs used for distillation.

\begin{figure*}[t]
    \centering
    \includegraphics[width=0.95\textwidth]{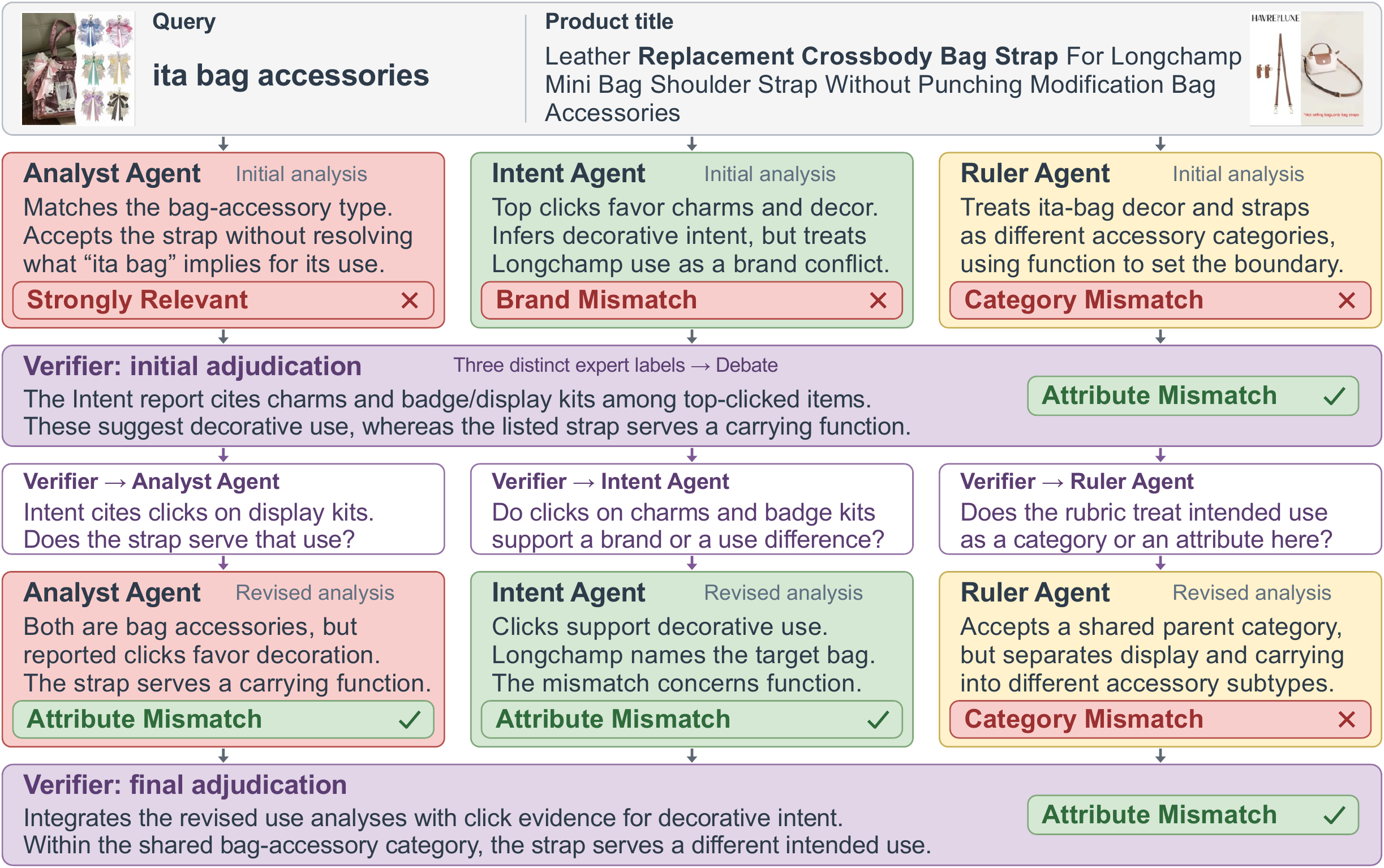}
    \caption{Debate refines expert analysis while preserving
    a correct adjudication.}
    \Description{A condensed and paraphrased interaction for
    ``ita bag accessories'' and a replacement bag strap.
    The Analyst, Intent, and Ruler initially predict
    Strongly Relevant, Brand Mismatch, and Category Mismatch,
    respectively, triggering debate.
    After focused questioning, the Analyst and Intent
    identify the difference between decorative and carrying
    uses and revise their labels to Attribute Mismatch.
    The Ruler retains Category Mismatch by applying
    a finer functional category boundary.
    The Verifier maintains Attribute Mismatch.}
    \label{fig:case_2}
\end{figure*}

\begin{figure*}[t]
    \centering
    \includegraphics[width=0.95\textwidth]{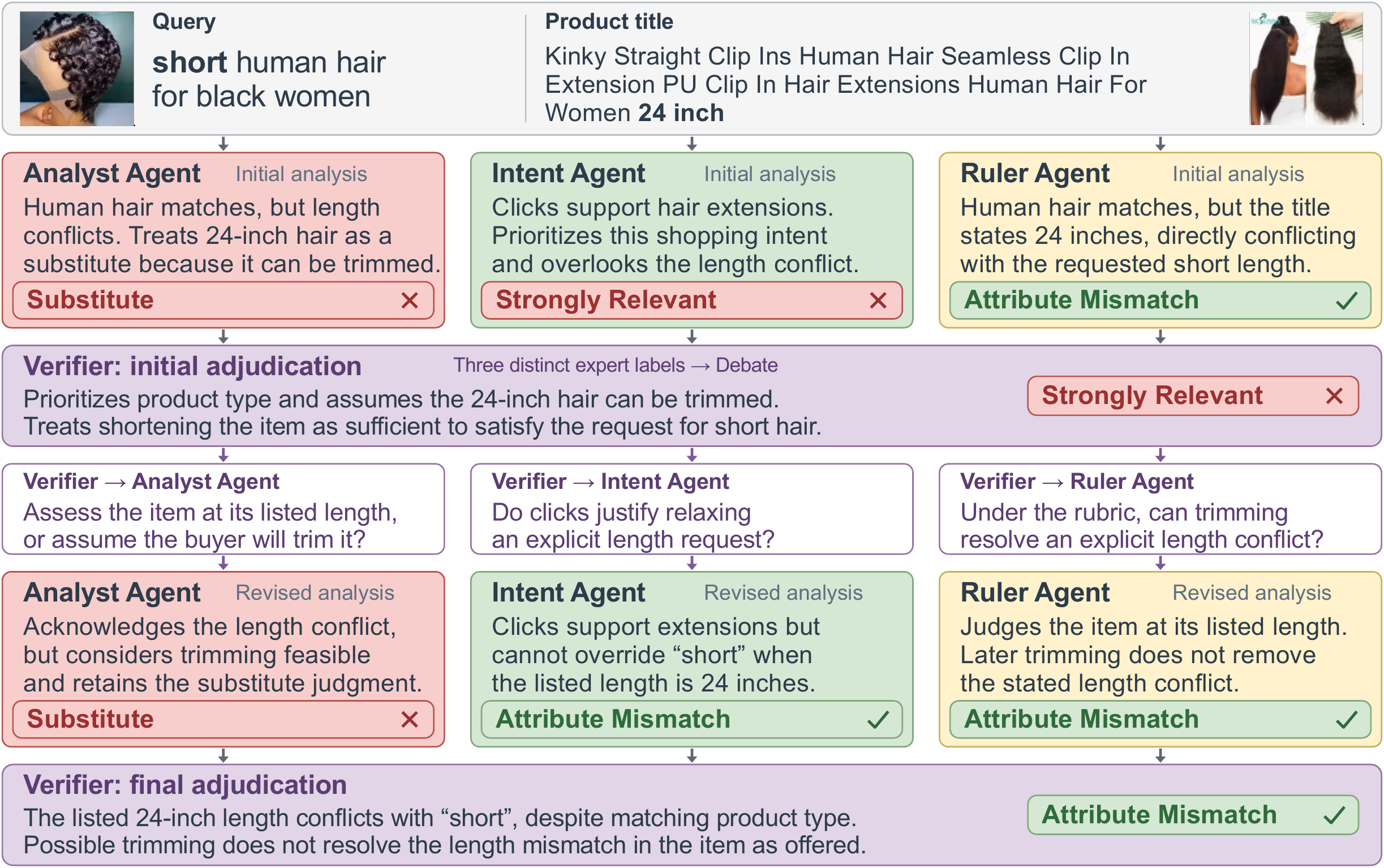}
    \caption{Debate corrects an initial Verifier error
    without expert consensus.}
    \Description{A condensed and paraphrased interaction for
    a short-hair query and human-hair extensions
    listed as 24 inches long.
    The Analyst predicts Substitute based on possible
    trimming, the Intent predicts Strongly Relevant,
    and the Ruler predicts Attribute Mismatch.
    These distinct labels trigger debate.
    The Analyst retains Substitute, while the Intent
    and Verifier revise to Attribute Mismatch.
    The Ruler maintains its judgment that the item
    should be assessed at its listed length.}
    \label{fig:case_3}
\end{figure*}

\subsection{Reward Implementation}
\label{app:reward-details}

This section specifies the scoring rules and
coefficients used in
Section~\ref{sec:composite-reward}.

\noindent\textbf{Adjudication quality.}
We compute
\begin{equation}
\begin{aligned}
r_{\mathrm{adj}} ={}&
0.1r_{\mathrm{fmt}}\\
&+\operatorname{Gate}_{\mathrm{fmt}}
\left(0.8r_{\mathrm{res}}
+0.1r_{\mathrm{reason}}\right),
\end{aligned}
\label{eq:adjudication-reward}
\end{equation}
where
$\operatorname{Gate}_{\mathrm{fmt}}
=r_{\mathrm{fmt}}\in\{0,1\}$
indicates whether the terminal prediction is
parseable as a valid relevance label. Let $\hat{y}$ denote the terminal prediction,
which is $\hat{y}_V^{(1)}$ without debate and
$\hat{y}_V^{(2)}$ otherwise, and let $y^*$
denote the gold label.
For a valid prediction, the outcome reward is
\begin{equation}
r_{\mathrm{res}}=
\begin{cases}
1, & \hat{y}=y^*,\\
0.65, & \hat{y}\neq y^*
        \ \text{and}\ \phi(\hat{y})=\phi(y^*),\\
0, & \text{otherwise},
\end{cases}
\label{eq:hierarchical-outcome}
\end{equation}
where $\phi$ maps each fine-grained label to
its relevant or irrelevant superclass.
Invalid terminal labels also receive
$r_{\mathrm{res}}=0$. The reasoning reward is
$r_{\mathrm{reason}}
=(c_{\mathrm{self}}+c_{\mathrm{struct}})/2$,
where $c_{\mathrm{self}}$  is computed by a deterministic rule that checks whether the label extracted from the rationale’s concluding step matches the terminal prediction, while $c_{\mathrm{struct}}$ checks
completion of the adjudication steps, coverage
of all three expert verdicts, and inclusion
of a concluding statement.

\noindent\textbf{Debate quality.}
When the three experts' first-round labels are
not unanimous, the debate reward is
\begin{equation}
r_{\mathrm{shape}}
=
b_{\mathrm{R1}}
+p_{\mathrm{fmt}}
+p_{\mathrm{leak}}
+s_{\mathrm{spec}}
+s_{\mathrm{shift}}.
\label{eq:debate-shaping}
\end{equation}
The first-round bonus $b_{\mathrm{R1}}$ is
$0.05$ when the initial Verifier judgment
matches the gold label and zero otherwise. Question quality is assessed through format
and leakage penalties together with a heuristic
specificity bonus.
A missing question or one whose length falls outside the permitted range incurs
$p_{\mathrm{fmt}}=-0.1$, while a focused question that matches a predefined regular-expression pattern for explicit disclosure of the Verifier’s own judgment incurs
$p_{\mathrm{leak}}=-0.1$.
Each penalty is zero when its corresponding
violation is absent.
The specificity bonus is
$s_{\mathrm{spec}}=0.05\rho$, where $\rho$
is the fraction of questions that exceed
60 characters and cite a query or product-title
token containing at least four characters.

For expert revisions, we compare each expert's predictions
before and after debate using
$c_k^{(t)}=\mathbf{1}[\hat{y}_k^{(t)}=y^*]$,
which indicates whether expert $k$ predicts the gold label
in round $t$.
The four cases are an incorrect prediction becoming correct,
a correct prediction remaining correct, an incorrect
prediction remaining incorrect, and a correct prediction
becoming incorrect.

The expert-revision reward is
\begin{equation}
\begin{aligned}
s_{\mathrm{shift}}
&=
\frac{0.1}{|\mathcal{K}|}
\sum_{k\in\mathcal{K}}\Delta_k,\\
\Delta_k
&=
\begin{cases}
1,   & (c_k^{(1)},c_k^{(2)})=(0,1),\\
0.3, & (c_k^{(1)},c_k^{(2)})=(1,1),\\
0,   & (c_k^{(1)},c_k^{(2)})=(0,0),\\
-1,  & (c_k^{(1)},c_k^{(2)})=(1,0).
\end{cases}
\end{aligned}
\label{eq:expert-revision}
\end{equation}

\section{Case Studies}
\label{app:case-studies}

Figures~\ref{fig:case_1}--\ref{fig:case_3} present
condensed and paraphrased representations of actual
model outputs on three test-set cases, illustrating
independent adjudication, expert refinement, and
Verifier correction.
The figures summarize the key evidence, judgments,
and interactions for readability, rather than
reproduce complete transcripts; full role-specific
input contexts are omitted for brevity.
\subsection{Correcting Unanimous Expert Errors}
\label{app:case-unanimous-error}

Figure~\ref{fig:case_1} pairs ``realme gt7 pro case''
with a case advertised for the GT7.
All three experts predict \textit{Strongly Relevant}, with
the Analyst overlooking ``Pro'' and the Intent and Ruler
Agents inferring compatibility from shopping intent and
a shared product line.
The Verifier checks these interpretations against the query
and title and rejects the unsupported compatibility assumption,
yielding \textit{Model Mismatch}.
Expert unanimity prevents debate but does not restrict the
Verifier's decision, illustrating independent adjudication
when all initial expert labels are incorrect.

\subsection{Refining Expert Analysis through Debate}
\label{app:case-divergent-errors}

Figure~\ref{fig:case_2} pairs ``ita bag accessories''
with a replacement strap for a Longchamp Mini Bag.
The Analyst predicts \textit{Strongly Relevant} from the
shared accessory category, the Intent Agent interprets
Longchamp as a brand conflict and predicts
\textit{Brand Mismatch}, and the Ruler assigns
\textit{Category Mismatch} based on accessory function.
The Verifier uses the Intent report's click evidence,
including charms and badge kits, to identify a
decorative need that the strap's carrying function does
not satisfy, yielding \textit{Attribute Mismatch}.
During debate, it relays this evidence to the Analyst
and asks the Intent Agent to distinguish a target-bag
reference from a brand requirement. Both revise to
\textit{Attribute Mismatch}, while the Ruler retains
\textit{Category Mismatch} under a finer functional boundary.
The Verifier maintains its correct judgment, illustrating
how the expert refinement rewarded by $s_{\mathrm{shift}}$
in Eq.~\eqref{eq:expert-revision} can improve supporting
analyses without changing the final label.

\subsection{Correcting Verifier Errors through Debate}
\label{app:case-debate-correction}

Figure~\ref{fig:case_3} pairs ``short human hair for black women''
with 24-inch hair extensions.
The Analyst predicts \textit{Substitute} based on possible
trimming, the Intent Agent prioritizes click-supported
interest in extensions and predicts \textit{Strongly Relevant},
and the Ruler predicts \textit{Attribute Mismatch} from
the explicit length conflict.
The Verifier initially accepts trimming and predicts
\textit{Strongly Relevant}.
During debate, its questions examine whether substitution
depends on modifying the item, whether behavioral evidence
justifies relaxing the length requirement, and how the
matching rubric treats hypothetical trimming.
The Analyst retains \textit{Substitute}, while the Intent
Agent revises to \textit{Attribute Mismatch} and the Ruler
reinforces assessment at the listed length.
The Verifier then recognizes that possible trimming does
not resolve the mismatch in the product as offered and
corrects its prediction to \textit{Attribute Mismatch},
despite persistent expert disagreement.
\end{document}